\documentclass[prl,twocolumn,showpacs,superscriptaddress,amssymb,floatfix]{revtex4}
\usepackage[dvips]{graphicx}
\usepackage{bm}
\newcommand{\Journal}[4]{#1 {\bf #2}, #3 (#4)}
\newcommand{\PR}{Phys. Rev.}
\newcommand{\PRL}{Phys. Rev. Lett.}
\newcommand{\PRA}{Phys. Rev. A}
\newcommand{\JMP}{J. Math. Phys.}
\newcommand{\EPJD}{Eur. Phys. J. D}
\newcommand{\Science}{Science}

\newcommand{\PLA}{Phys. Lett. A}
\newcommand{\NJP}{New Journal of Physics}
\begin{document}
\title {Fermi-Bose mapping and 
N-particle ground state of spin-polarized fermions in tight atom waveguides}
\author{M. D. Girardeau}
\email{girardeau@optics.arizona.edu}
\affiliation{Optical Sciences Center, University of Arizona, Tucson, AZ 85721}
\author{M. Olshanii}
\email{olshanii@phys4adm.usc.edu}
\affiliation{Department of Physics and Astronomy, University of Southern 
California, Los Angeles, CA 90089-0484}
\date{\today}
\begin{abstract}

A K-matrix for wave-guide confined spin-polarized fermionic atoms recently 
computed by Granger and Blume is identified, in the low-energy domain, 
with a contact condition for one-dimensional (1D) spinless fermions. 
Difficulties in consistently formulating the contact conditions in terms
of interaction potentials are discussed 
and a rigorous alternative variational reformulation is constructed.
A duality between 1D fermions and bosons with zero-range interactions 
suggested by Cheon and Shigehara is shown to hold for the effective 
1D dynamics of a spin-polarized Fermi gas with 3D p-wave 
interactions and that of a Bose gas with 3D s-wave interactions in a tight 
waveguide. This generalizes the mapping
from impenetrable bosons (TG gas) to free fermions and is used to derive 
the equation of state of an ultracold spin-polarized fermionic vapor in a 
tight waveguide. Near a 1D confinement-induced resonance one has a 
``fermionic TG gas'' which maps to an \emph{ideal} Bose gas.
\end{abstract}
\maketitle
Ultracold atomic vapors in atom waveguides are currently
a subject of great experimental and theoretical interest and activity
due to potential applicability to atom interferometry \cite{Ber97,Dum02} 
and integrated atom optics \cite{Dek00,Sch98} and their utility
for demonstrating novel highly-correlated quantum states. 
Exploration of these systems is facilitated by tunability of their
interactions by external magnetic fields via Feshbach
resonances \cite{Rob01}. In fermionic atoms in the same spin state,
s-wave scattering is forbidden by the exclusion principle and p-wave 
interactions are usually 
negligible. However, they can be greatly enhanced by Feshbach resonances,
which have recently been observed in an ultracold atomic vapor of
spin-polarized fermions \cite{RegTicBohJin03}. Additional resonances 
are induced by tight transverse confinement in an atom waveguide. Of
particular interest is the regime of low temperatures
and densities where transverse oscillator modes are frozen and the dynamics 
is described by an effective 1D Hamiltonian with zero-range interactions
\cite{Ols98,PetShlWal00}, a regime already reached 
experimentally \cite{Bon01,Gre02,GorVogLea01}. Transverse modes are still 
\emph{virtually} excited during
collisions, leading to renormalization of the effective 1D coupling
constant $g_{1D}$ via a confinement-induced resonance. This was first 
shown for bosons \cite{Ols98} and recently explained in terms 
of Feshbach resonances associated with bound states in closed, virtually 
excited transverse oscillation channels \cite{BerMooOls03}. Recently
the analogous problem for fermions has been solved by Granger 
and Blume \cite{GraBlu03}, who have shown that such 
resonances also occur in spin-polarized fermionic vapors. Investigation
of such systems is facilitated by a mapping which allows
reduction of strongly interacting fermions in one dimension to
weakly interacting bosons.
An energy-dependent mapping of this type
was demonstrated in this recent work \cite{GraBlu03}. The analysis herein
will be limited to a low-energy regime where one can use a simpler 
mapping originally employed to reduce 
the 1D hard core Bose gas to an ideal Fermi gas 
\cite{Gir60,Gir65}. Here, following Cheon and Shigehara \cite{CheShi98}, 
the same mapping will be employed to map the 
\emph{strongly interacting} Fermi gas to a \emph{weakly interacting}
Bose gas. More generally, for all values of the effective 1D fermionic
coupling constant, the known ground state of the
1D Bose gas with delta-function repulsion, the Lieb-Liniger (LL) gas
\cite{LieLin63}, will be mapped to generate the 1D
ground state of a spin-polarized Fermi gas with zero-range
p-wave interactions. 

{\it Contact condition for spin-polarized fermions in a waveguide:}
Granger and Blume derived the effective one-dimensional K-matrix for 
two interacting fermions 
confined in a single-mode harmonic atom waveguide \cite{GraBlu03}. 
It can be shown that in the low-energy \cite{Note0} domain the K-matrix 
can be reproduced, with a relative error as small as ${\cal O}(k_{z}^3)$, 
by the contact condition
\begin{equation}\label{Fermi-contact}
\psi_{F}(0+)=-\psi_{F}(0-)= -a_{1D}^{F}\psi_{F}^{'}(0\pm)
\end{equation}
where 
\begin{eqnarray}\label{renorm}
a_{1D}^{F}&=&\frac{6V_{p}}{a_{\perp}^2}[1+12(V_{p}/a_{\perp}^3)
|\zeta(-1/2,1)|]^{-1} 
\end{eqnarray}
is the odd-wave one-dimensional scattering length, 
$V_{p}=a_{p}^{3}=-\lim_{k\to 0}\tan\delta_{p}(k)/k^3$ is the 
p-wave ``scattering volume'' \cite{SunEsrGre03}, $a_{p}$ is the p-wave
scattering length, $a_{\perp}=\sqrt{\hbar/\mu\omega_{\perp}}$ is the 
transverse oscillator length \cite{Note1},
$\zeta(-1/2,1)=-\zeta(3/2)/4\pi=-0.2079\ldots$ is 
the Hurwitz zeta function evaluated at 
$(-1/2,1)$ \cite{WhiWat52}, and $\mu$ is the reduced mass. 
The expression (\ref{renorm}) has a resonance
at a \emph{negative} critical value  
$V_{p}^{crit}/a_{\perp}^{3}=-0.4009\cdots$.
In accordance with (\ref{Fermi-contact}),  
the low-energy fermionic wavefunctions, Eq. (20)
of \cite{GraBlu03}, are discontinuous at 
contact, but left and right limits of their derivatives coincide.
Following \cite{CheShi98} we assume the same here. 

{\it Odd-wave  one-dimensional interaction potential:} 
Following the even-wave (bosonic) case, where 
the $\delta$-interaction can be introduced naturally to cancel 
the $\delta$-functions resulting from double-differentiation of functions with 
discontinuous derivatives, in the case of fermions whose wave function is 
discontinuous 
it is tempting to introduce $\delta'$ interactions. However, 
$\delta'$-functions and second derivatives 
are known to be ill-defined if used in a convolution with 
discontinuous functions, making a consistent Hamiltonian formulation and
corresponding perturbative treatments impossible. However, a consistent 
{\it variational} formulation 
does exist, where matrix elements of operators are 
replaced by {\it two-slot} functionals not factorizable as standard 
``bra-operator-ket'' products. Such a formulation does allow 
an accurate first order perturbation theory, and higher  
orders are under investigation. Furthermore, an exact Fermi-Bose mapping to be
discussed allows nonperturbative treatment in the equivalent bosonic space. 

Consider general contact conditions 
\begin{eqnarray}\label{General-contact}
\psi'(0+)-\psi'(0-)= -(a_{1D}^{B})^{-1}[\psi(0+) + \psi(0-)]\nonumber
\\
\psi(0+)-\psi(0-)= -a_{1D}^{F}[\psi'(0+) + \psi'(0-)]
\end{eqnarray}
that can scatter both even and odd partial waves. Here $a_{1D}^{B}$
($a_{1D}^{F}$) is the even (odd) scattering length. Our goal is 
to identify a functional whose extrema are solutions of the 
free-space Schr\"{o}dinger equation subject to the contact conditions 
(\ref{General-contact}). Introduce ``two-slot'' Hermitian functionals 
corresponding to the ``square-derivative'' and $\delta$-function respectively:
\begin{eqnarray}\label{two-slot}
&&\int dz\, {\chi^{*}}'\, \psi'\
\equiv \left(\int^{0-} + \int_{0+}\right) dz\,
{\chi^{*}}'\, \psi'\
\nonumber
\\
&&\quad + \frac{1}{2} [\chi^{*}(0+)-\chi^{*}(0-)][\psi'(0+)+\psi'(0-)]
\nonumber
\\
&&\quad + \frac{1}{2} [{\chi^{*}}'(0+)+{\chi^{*}}'(0-)][\psi(0+)-\psi(0-)]
\label{functionals}
\\
&&\int dz\, \chi^{*}\, \delta(z) \, \psi
\equiv
\frac{1}{4} [\chi^{*}(0+)+\chi^{*}(0-)][\psi(0+)+\psi(0-)]\ .\nonumber
\end{eqnarray}
After a lengthy but straightforward calculation one can show
that extrema $\psi$ of the energy functional
\begin{eqnarray}
&&{\cal E} = \hbar^2/2\mu \int dz\, {\psi^{*}}'\,\psi'
+g_{1D}^{B} \int dz\, \psi^{*}\delta(z) \, \psi
\nonumber
\\
&&\quad +g_{1D}^{F} \int dz\, {\psi^{*}}'\delta(z) \, \psi'
\end{eqnarray}
with integrals defined by (\ref{two-slot}) with $\chi=\psi$ 
and the variational space spanned by (normalized) wave functions 
with {\it arbitrary} discontinuities at zero, do obey the 
contact conditions (\ref{General-contact}), 
being local eigenstates of the kinetic energy outside of the 
contact point $z=0$. Here the coupling constants are given by 
$g_{1D}^{B} = -\hbar^{2}/\mu a_{1D}^B$ and 
$g_{1D}^{F} = +\hbar^{2}a_{1D}^{F}/\mu$. 
One may introduce a formal ``Hamiltonian'' for the relative motion of two 
fermions by $\hat{H}_{1D}^{F}=(\hbar^{2}/2\mu) 
{^\leftarrow}(\partial_{z}) (\partial_{z})^\rightarrow
+g_{1D}^{F}\, {^\leftarrow}(\partial_{z}) \, 
\delta(z)\,(\partial_{z})^\rightarrow$ 
where this ``operator'' (and especially the kinetic energy part of it) 
must never appear outside of matrix elements, which 
should be carefully computed using the rules 
(\ref{functionals}), and the eigenvalue problem for 
this Hamiltonian must be replaced by a variational one.
Notice that according to (\ref{functionals}), the kinetic energy operator
$(\hbar^{2}/2\mu) 
{^\leftarrow}(\partial_{z}) (\partial_{z})^\rightarrow$ is a 
``regularized kinetic
energy'' defined in such a way that the product of two $\delta$ function
contributions is automatically subtracted from the result of insertion
of this operator between two discontinuous functions. 
We have verified that for two fermions in an anti-periodic box the 
``potential'' $g_{1D}^{F} \, {^\leftarrow}(\partial_{z}) \, 
\delta(z)\,(\partial_{z})^\rightarrow$ correctly reproduces the 
first order perturbation theory correction to the energy, but we 
warn the reader that the formal similarity 
between functionals (\ref{functionals}) and matrix elements of real 
operators should \emph{not} lead to an attempt to reformulate the problem 
as a matrix diagonalization with some basis set. For example, one can 
check that in momentum space such a procedure leads to ultraviolet 
divergences, and an attempt to cancel them leads again to an unfactorizable 
functional. However, the contact conditions plus the \emph{free-particle} 
Schr\"{o}dinger equation for $z\ne 0$ \emph{do} define a well-posed 
eigenvalue problem not requiring use of a formal interaction operator.

{\it Fermi-Bose mapping:} On the space of antisymmetric functions 
$\psi_F$ the contact conditions (\ref{General-contact}) reduce to
$\psi_{F}(0+)=-\psi_{F}(0-)=-a_{1D}^{F}\psi_{F}^{'}(0\pm)$ with
$\psi_{F}^{'}(0+)=\psi_{F}^{'}(0-)$, and on the space of symmetric 
functions $\psi_B$ they reduce to 
$\psi_{B}^{'}(0+)=-\psi_{B}^{'}(0-)=-(a_{1D}^{B})^{-1}\psi_{B}(0\pm)$ 
with $\psi_{B}(0+)=\psi_{B}(0-)$. Defining symmetric wave
functions $\psi_{B}(z)=\text{sgn}(z)\psi_{F}(z)$ and mapped scattering length
$a_{1D}^{B}=a_{1D}^{F}\equiv a_{1D}$ 
where $\text{sgn}(z)$ is $+1$ if $z>0$ and $-1$ if $z<0$, one finds that
the Bose and Fermi contact conditions are equivalent. Since the kinetic
energy contributions from $z\ne 0$ also agree, one has a mapping from
the fermionic to bosonic problem which preserves energy 
eigenvalues and dynamics. The relation between coupling constants
$g_{1D}^{F}$ in $\hat{H}_{1D}^{F}$ and $g_{1D}^{B}$ in
$\hat{H}_{1D}^{B}=-(\hbar^{2}/2\mu)\partial_{z}^{2}+g_{1D}^{B}\delta(z)$
is $g_{1D}^{B}=-\hbar^{4}/\mu^{2}g_{1D}^{F}$, and by
(\ref{renorm}) this agrees with the low-energy
limit of Eq. (25) of \cite{GraBlu03,Note0}. 
In the limit $g_{1D}^{B}=+\infty$ arising when $V_{p}\to 0-$, this is
the $N=2$ case of the original mapping \cite{Gir60,Gir65} from hard sphere 
bosons to an ideal Fermi gas, but now generalized to 
arbitrary
coupling constants and used in the inverse direction. This generalizes to
arbitrary $N$: Fermionic solutions $\psi_{F}(z_{1},\cdots,z_{N};t)$
are mapped to bosonic solutions $\psi_{B}(z_{1},\cdots,z_{N};t)$ via
$\psi_{B}=A(z_{1},\cdots,z_{N})\psi_{F}(z_{1},\cdots,z_{N};t)\psi_F$
where $A=\prod_{1\le j<\ell\le N}\text{sgn}(z_{j\ell})$
is the same mapping function used originally \cite{Gir60,Gir65}. 
The Fermi contact conditions are 
$\psi_{F}|_{z_{j}=z_{\ell}+}=-\psi_{F}|_{z_{j}=z_{\ell}-}
=-(a_{1D}/2)(\partial_{z_{j}}-\partial_{z_{\ell}})
\psi_{F}|_{z_{j}=z_{\ell\pm}}$ and imply the Bose contact conditions
$(\partial_{z_{j}}-\partial_{z_{\ell}})\psi_{B}|_{z_{j}=z_{\ell}+}
=-(\partial_{z_{j}}-\partial_{z_{\ell}})\psi_{B}|_{z_{j}=z_{\ell}-}
=-(2/a_{1D})\psi_{B}|_{z_{j}=z_{\ell}}$ with  
$a_{1D}\equiv a_{1D}^{B}=a_{1D}^{F}$, and these are the
usual LL contact conditions \cite{LieLin63}.
This mapping remains valid if external potentials $v_{ext}(z_j)$
and/or additional interactions $v_{l.r.}(z_{j\ell})$ of nonzero range
are present. One can define a formal fermionic Hamiltonian
$\hat{H}_{1D}^F=(\hbar^{2}/2\mu)\sum_{j=1}^{N}
{^\leftarrow}(\partial_{z_j}) (\partial_{z_j})^\rightarrow
+g_{1D}^{F}\sum_{1\le j<\ell\le N}{^\leftarrow}(\partial_{z_{j\ell}}) \, 
\delta(z)\,(\partial{z_{j\ell}})^\rightarrow$, 
but we again warn the reader that in calculations it should be treated
variationally or its ``interaction'' term replaced
by the contact conditions, and it must \emph{not} be
substituted into a second-quantized framework. On the other hand, 
the definition of the interaction term in 
$\hat{H}_{1D}^{B}=-(\hbar^2/2\mu)\sum_{j=1}^{N}\partial_{z_j}^{2}
+g_{1D}^{B}\sum_{1\le j<\ell\le N}\delta(z_{j\ell})$ is much less
delicate, so calculations, including
second quantization if desired, can be performed in the 
mapped bosonic Hilbert space. 

\begin{figure}
\includegraphics[width=1.0\columnwidth,angle=0]{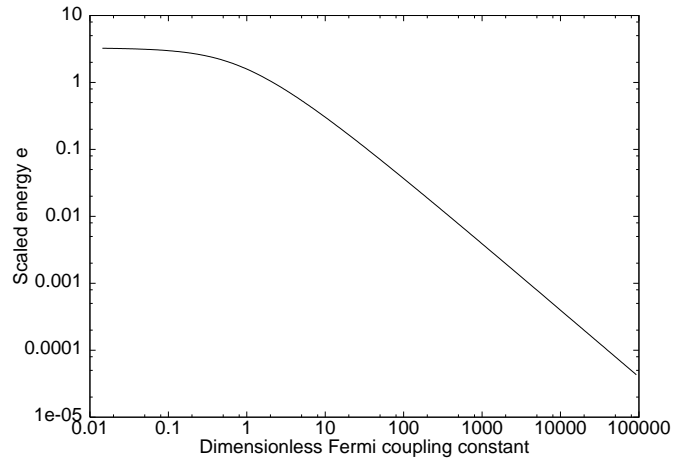}
\caption{Log-log plot of scaled ground state energy per particle 
$e=2m\epsilon/\hbar^{2}n^2$
versus dimensionless fermionic coupling constant $\gamma_F$.}
\label{Fig:one}
\vspace{-0.5cm}
\end{figure}

\emph{N-particle states:} The exact ground
\cite{LieLin63} and excited \cite{Lieb63} states of $\hat{H}_{1D}^B$
are known for all positive
$g_{1D}^B$ if no external potential or nonzero range interactions
are present, and the mapping then generates the exact $N$-body ground
and excited states of $\hat{H}_{1D}^F$. Define
dimensionless bosonic and fermionic coupling constants by 
$\gamma_{B}=mg_{1D}^{B}/n\hbar^2$ and 
$\gamma_{F}=-mg_{1D}^{F}n/\hbar^2$ where $n$ is the longitudinal
particle number density and the minus prefactor of $\gamma_F$ is convenient 
since $g_{1D}^B$ and $g_{1D}^F$ have opposite signs. They satisfy 
$\gamma_{B}\gamma_{F}=4$. The ground state energy per particle $\epsilon$ is
related to a dimensionless function $e(\gamma)$ available online
\cite{Note4} via $\epsilon=(\hbar^{2}/2m)n^{2}e(\gamma)$ where $\gamma$
is related to $\gamma_F$ herein by $\gamma=\gamma_{B}=4/\gamma_{F}$.
This is plotted as a function of $\gamma_F$ in Fig. \ref{Fig:one}.
If $\gamma_F$ and $\gamma_B$ are negative the Bose gas and mapped Fermi gas
are unstable against collapse to an ultrahigh density droplet (``bright
soliton'') in the absence of longitudinal trapping \cite{McG64}. 
However, the derivation of an effective 1D Hamiltonian from the 3D one
breaks down in the collapsed regime, as does neglect of three-particle and
in fact multiparticle interatomic interactions. In the case 
of longitudinal trapping the gaseous regime is probably 
metastable if $|\gamma_{B}|$ is not too large (hence $|\gamma_{F}|$ not 
too \emph{small}), but this is beyond the scope of the present treatment.

\begin{figure}
\includegraphics[width=1.0\columnwidth,angle=0]{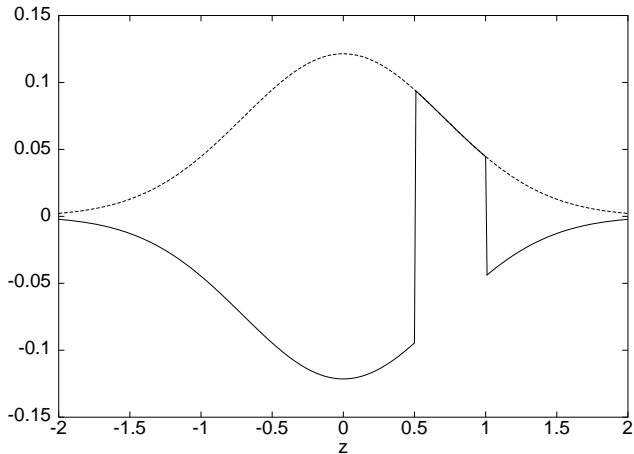}
\caption{$\psi_{0F}(z,z_{2},z_{3})$ (solid line) and 
$\psi_{0B}(z,z_{2},z_{3})$ (dashed line) for a longitudinally trapped 
fermionic TG gas, as a function of $z$ for $z_{2}=0.5$ and $z_{3}=1$.
Units are such that $a_{long}=1$.}
\label{Fig:two}
\vspace{-0.5cm}
\end{figure}

\begin{figure}
\includegraphics[width=1.0\columnwidth,angle=0]{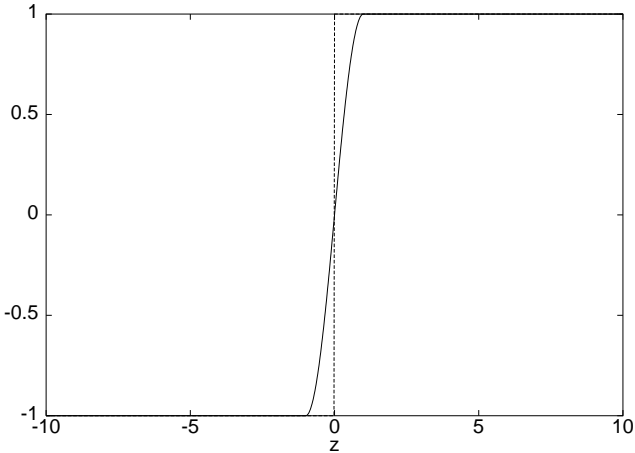}
\caption{$N=2$ untrapped fermionic TG gas ground state (dashed line) 
compared with zero-energy scattering solution for a square well 
with range $z_0$ and depth $V_0$ corresponding to the boundary between no 
bound state and one bound state, a zero energy resonance (solid line), 
as function of relative coordinate z. Units are such that $z_0=1$.} 
\label{Fig:three}
\vspace{-0.5cm}
\end{figure}

\emph{Fermionic TG gas:} The mapping $\psi_{B}=A\psi_F$ was originally 
introduced to map the strongly-interacting many-body problem of 
1D hard-sphere bosons of diameter $d$ to the ideal Fermi gas 
\cite{Gir60,Gir65}. The simplest case $d\to 0+$, the impenetrable point
1D Bose gas, has recently elicited a great deal of theoretical 
\cite{GirWri00,GirWriTri01,DasLapWri02,DasGirWri02,GirDasWri02}
and experimental \cite{GorVogLea01,Gre02} 
activity in the context of bosonic atomic vapors in tight atom wave guides, 
where it is now called the Tonks-Girardeau (TG) gas 
\cite{DunLorOls01,OlsDun02,OhbSan02,GanShl03,AstGio02,
BogMalBulTim03,LieSeiYng03}. For bosons the TG regime is reached when
$g_{1D}^{B}$ is large enough and/or the density $n$ \emph{low} enough that 
$\gamma_{B}\gg 1$. A similar simplification occurs in the fermionic
case, where a fermionic TG regime is reached when $g_{1D}^{F}$ is 
\emph{negative} and
large enough and/or $n$ \emph{high} enough that $\gamma_{F}\gg 1$.
The corresponding fermionic TG gas then maps to the \emph{ideal} Bose gas
since $\gamma_{B}\gamma_{F}=4$. As an example, suppose that there is
a longitudinal trap potential $v_{ext}(z)=(m/2)\omega_{long}^{2}z^2$.
Then in the fermionic TG limit the $N$-boson ground state is
$\psi_{B}(z_{1},\cdots,z_{N})=\prod_{j=1}^{N}u_{0}(z_{z})$ with 
$u_{0}(z)=\pi^{-1/4}a_{long}^{-1/2}e^{-(z/a_{long})^{2}}$ with
$a_{long}=\sqrt{\hbar/\mu\omega_{long}}$, and the corresponding fermionic
TG ground state $\psi_{F}=A\psi_B$ has discontinuities
at collisions $z_{j}=z_{\ell}$. Fig. \ref{Fig:two} shows $\psi_{0}^F$
and $\psi_{0}^B$ for $N=3$. The discontinuities in $\psi_{0}^F$
are a consequence of idealization to a zero-range pseudopotential.
For a potential of nonzero range $r_{0}\ll a_p$ they 
are rounded over a distance $\ll a_p$. 
As an illustration, Fig. \ref{Fig:three} compares the two-particle
ground state of the untrapped fermionic TG gas with the solution when
the zero-range interaction is replaced by a square well
potential equal to $-V_0$ when $-z_{0}<z<z_0$ and zero when $|z|>z_0$.
(Note that the interaction term in $\hat{H}_{1D}^F$ is \emph{negative}
definite in the regime of interest, where $g_{1D}^{F}<0$ and 
$g_{1D}^B>0.$ )
The energy is taken as zero so the exterior solution is 
$\text{sgn}(z)=\pm 1$; an interior
solution fitting smoothly onto this is $\sin(\kappa z)$ with
$\kappa=\sqrt{2\mu V_{0}/\hbar^2}=\pi/2z_0$, the critical
value where the last bound state passes into the continuum, 
a zero-energy resonance. A fermionic contact condition with a finite
scattering length can be obtained in the limit $z_{0}\to 0$ if $\kappa$
scales with the width $z_{0}$ as 
$\kappa=(\pi/2z_{0})[1+(2/\pi)^2 (z_{0}/a_{1D}^{F})]$.

\emph{Discussion:} The effective 1D $N$-particle ground state of a
spin-polarized Fermi gas with zero-range p-wave interactions 
has been mapped to the $N$-particle
ground state of the 1D Bose gas with delta-function repulsion
\cite{LieLin63}, providing the exact solution of this
fermionic problem in the absence of longitudinal trapping. Experiments on 
spin-polarized Fermi gases in this quasi-1D regime are suggested, as is 
investigation of Fermi-Bose duality for 
waveguide-confined ultracold gases with realistic interactions. 
\begin{acknowledgments}
We are very grateful to 
Doerte Blume for communications regarding her closely-related
work with Brian Granger \cite{GraBlu03} and a draft of subsequent work 
\cite{AstBluGioGra03}.
This work was supported by Office of Naval Research grant N00014-03-1-0427
(M.D.G. and M.O.) and by NSF grant PHY-0301052 (M.O.).
\end{acknowledgments}

\begin{thebibliography}{39}
%
\bibitem{Ber97} P.R. Berman, ed., \emph{Atom Interferometry} (Academic
Press, Boston, 1997).
%
\bibitem{Dum02} R. Dumke {\it et al.}, \Journal{\PRL}{89}{220402}{2002}.
%
\bibitem{Dek00} N.H. Dekker {\it et al.}, \Journal{\PRL}{84}{1124}{2000}.
%
\bibitem{Sch98} J. Schmiedmayer, \Journal{\EPJD}{4}{57}{1998}.
%
\bibitem{Rob01} J.L. Roberts, N.R. Claussen, S.L. Cornish, E.A. Donley,
E.A. Cornell, and C.E. Wieman, \Journal{\PRL}{86}{4211}{2001}.
%
\bibitem{RegTicBohJin03} C.A. Regal, C. Ticknor, J.L. Bohn, and D.S. Jin,
\Journal{\PRL}{90}{053201}{2003}.
%
\bibitem{Ols98} M. Olshanii, \Journal{\PRL}{81}{938}{1998}.
%
\bibitem{PetShlWal00} D.S. Petrov, G.V. Shlyapnikov, and J.T.M. Walraven, 
\Journal{\PRL}{85}{3745}{2000}.
%
\bibitem{Bon01} K. Bongs {\it et al.}, \Journal{\PRA}{63}{031602}{2001}.
%
\bibitem{Gre02} M. Greiner {\it et al.}, \Journal{\PRL}{87}{160405}{2001}
and Nature {\bf 415}, 39 (2002).
%
\bibitem{GorVogLea01} A. G\"{o}rlitz {\it et al}, \Journal{\PRL}{87}{130402}{2001}. 
%
\bibitem{BerMooOls03} T. Bergeman, M. Moore, and M. Olshanii,
arXiv:cond-mat/0210556 v2 (2003).
%
\bibitem{GraBlu03} B.E. Granger and D. Blume, arXiv:cond-mat/0307358 (2003).
%
\bibitem{Gir60} M. Girardeau, \Journal{\JMP}{1}{516}{1960}. For the case
$d>0$ see footnote 7(a).
%
\bibitem{Gir65} M.D. Girardeau, \Journal{\PR}{139}{B500}{1965}, 
particularly Secs. 2, 3, and 6.
%
\bibitem{CheShi98} T. Cheon and T. Shigehara, \Journal{\PLA}{243}{111}{1998}
and \Journal{\PRL}{82}{2536}{1999}.
%
\bibitem{LieLin63} Elliott H. Lieb and Werner Liniger, Phys. Rev. {\bf 130}, 
1605 (1963).
%
\bibitem{Note0} Their $E$ includes transverse zero-point energy
$\hbar\omega_{\perp}$, so zero
longitudinal energy implies $E=\hbar\omega_{\perp}$.
%
\bibitem{SunEsrGre03} H. Suno, B.D. Esry, and C.H. Greene,
\Journal{\PRL}{90}{053202}{2003}.
%
\bibitem{Note1} The oscillator length is often defined as 
$\sqrt{\hbar/m\omega_{\perp}}$. Here the effective mass $\mu$ is used instead
of $m$, for ready comparison with recent literature
\cite{Ols98,BerMooOls03,GraBlu03}.
%
\bibitem{WhiWat52} E.T. Whittaker and G.N. Watson, \emph{A Course of
Modern Analysis} (Cambridge University Press, 1952), pp. 265-269,
particularly the top equation on p. 269 for negative first argument.
%
%\bibitem{MooBerOls03} T. Bergeman, M. Moore, and M. Olshanii, to be published.
%
\bibitem{Lieb63} Elliott H. Lieb, \Journal{\PR}{130}{1616}{1963}.
%
\bibitem{Note4} $e(\gamma)$ is made available online by V. Dunjko and
M. Olshanii at 
http://physics.usc.edu/$\widetilde{\hspace{0.15cm}}$olshanii/DIST/.
% 
\bibitem{McG64} J.B. McGuire, \Journal{\JMP}{5}{622}{1964}, Sec. VI.C.
%
\bibitem{GirWri00} M.D. Girardeau and E.M. Wright,
\Journal{\PRL}{84}{5691}{2000} and \Journal{\PRL}{84}{5239}{2000}.
%
\bibitem{GirWriTri01} M.D. Girardeau, E.M. Wright, and J.M. Triscari,
\Journal{\PRA}{63}{033601}{2001}.
%
\bibitem{DasLapWri02} Kunal K. Das, G. John Lapeyre, and E.M. Wright, 
Phys. Rev. A {\bf 65}, 063603 (2002).
%
\bibitem{DasGirWri02} Kunal K. Das, M.D. Girardeau, and E.M. Wright, 
Phys. Rev. Lett. {\bf 89}, 170404 (2002).
%
\bibitem{GirDasWri02}
M.D. Girardeau, Kunal K. Das, and E.M. Wright,
Phys. Rev. A {\bf 66}, 023604 (2002)
%
\bibitem{DunLorOls01} V. Dunjko, V. Lorent, and M. Olshanii, 
\Journal{\PRL}{86}{5413}{2001}.
%
\bibitem{OlsDun02} M. Olshanii and V. Dunjko, \Journal{\PRL}{91}{090401}
{2003}.
%
\bibitem{OhbSan02} P. \"{O}hberg and L. Santos, \Journal{\PRL}{89}{240402}
{2002}.
%
\bibitem{GanShl03} D.M. Gangardt and G.V. Shlyapnikov, 
\Journal{\PRL}{90}{010401}{2003} and \Journal{\NJP}{5}{79.1}{2003}.
%
\bibitem{AstGio02} G.E. Astrakharchik and S. Giorgini, 
arXiv:cond-mat/0212512 (2002).
%
\bibitem{BogMalBulTim03} N.M. Bogoliubov, C. Malyshev, R.K. Bullough,
and J. Timonen, arXiv:cond-mat/0306735 (2003).
%
\bibitem{LieSeiYng03} E.H. Lieb, R. Seiringer, and J. Yngvason,
arXiv:cond-mat/0304071 (2003).
%
\bibitem{OHaHemGehGraTho02} K.M. O'Hara, S.L. Hemmer, M.E. Gehm,
S.R. Granade, and J.E. Thomas, \Journal{\Science}{298}{2179}{2002}.
%
\bibitem{Bou03} T. Bourdel {\it et al.}, \Journal{\PRL}{91}{020402}{2003}. 
%
\bibitem{AstBluGioGra03} G.E. Astrakharchik, D. Blume, S. Giorgini,
and B.E. Granger, arXiv:cond-mat/0308585 (2003).
%
\end{thebibliography}
\end{document}